# Enhancement of critical temperature in fractal metamaterial superconductors

Igor I. Smolyaninov [1)] and Vera N. Smolyaninova [2)]

[1]Department of Electrical and Computer Engineering, University of Maryland, College Park, MD 20742, USA

[2]Department of Physics Astronomy and Geosciences, Towson University, 8000 York Road, Towson, MD 21252, USA

**Fractal metamaterial superconductor geometry has been suggested and analyzed based on the recently developed theoretical description of critical temperature increase in epsilon near zero (ENZ) metamaterial superconductors. Considerable enhancement of critical temperature has been predicted in such materials due to appearance of large number of additional poles in the inverse dielectric response function of the fractal. Our results agree with the recent observation (Fratini et al. Nature 466, 841 (2010)) that fractal defect structure promotes superconductivity.**

Metamaterials are artificially designed materials, which derive their properties from their engineered structure rather than their chemical composition. While searching for natural materials exhibiting higher critical temperatures constitutes a traditional approach to high temperature superconductivity research, metamaterial approach consists in artificial engineering of the superconductor in such a way that the electron pairing interactions in the metamaterial is maximized by its structure. Viability of this alternative approach has been illustrated in our recent theoretical [1-3] and experimental [4-6] work, which demonstrated that many tools developed in electromagnetic metamaterial research may be successfully used to engineer artificial metamaterial superconductors having considerably improved superconducting properties. This unexpected connection between electromagnetic metamaterials and superconductors points out to the link between superconducting properties and the effective dielectric response function $\varepsilon_{eff}(q,\omega)$ [7]. Therefore, it is natural to use our ability to engineer the dielectric response function $\varepsilon_{eff}(q,\omega)$ of the metamaterial to maximize the electron pairing interaction. Theoretical and experimental studies of such well-known metamaterial geometries as hyperbolic [8] and epsilon near zero (ENZ) [9] metamaterials based on either random nanoparticle mixtures [4] or metal-dielectric core-shell nanoparticles [5] indeed demonstrated considerable enhancement of superconducting properties compared to their parent materials. The most striking example of successful metamaterial superconductor engineering was recent observation of tripling of the critical temperature $T_c$ in Al-Al$_2$O$_3$ core-shell superconductors [5]. This success is even more compelling in light of the recently developed theoretical model of such an ENZ metamaterial, which predicts its $T_c$ enhancement based on the known superconducting properties of bulk aluminum [3]. In this paper we capitalize on this success and suggest and analyze novel fractal metamaterial superconductor geometry. Considerable enhancement of critical temperature is predicted in such fractal metamaterials due to appearance of a large number of additional poles in the inverse dielectric response function of the fractal. Our results also agree with the recent observations [10, 11] that fractal defect structure promotes superconductivity.

First, let us briefly state the basic features of our model of critical temperature increase in metal-dielectric ENZ metamaterials [3]. Within the framework of macroscopic electrodynamics, the electron pairing interaction in a superconductor may be expressed in the form of an effective Coulomb potential

$$V(\vec{q},\omega) = \frac{4\pi e^2}{q^2 \varepsilon_{\it eff}(\vec{q},\omega)}, \qquad (1)$$

where $V_C = 4\pi e^2/q^2$ is the Fourier-transformed Coulomb potential in vacuum, and $\varepsilon_{\it eff}(q,\omega)$ is the linear dielectric response function of the superconductor treated as an effective medium. Such a macroscopic electrodynamics description is valid if the material may be considered as a homogeneous medium on the spatial scales below the superconducting coherence length [7]. The critical temperature of a superconductor in the weak coupling limit is calculated as

$$T_c = \theta \ \exp\left(-\frac{1}{\lambda_{\it eff}}\right), \qquad (2)$$

[12] where $\theta$ is the characteristic temperature for a bosonic mode mediating electron pairing (such as the Debye temperature $\theta_D$ in the standard BCS theory), and $\lambda_{\it eff}$ is the dimensionless coupling constant defined by $V(q,\omega) = V_C(q)\,\varepsilon^{-1}(q,\omega)$ and the density of states $\nu$ (see for example [13]):

$$\lambda_{\it eff} = -\frac{2}{\pi}\nu \int_0^\infty \frac{d\omega}{\omega}\langle V_C(q)\,{\rm Im}\,\varepsilon^{-1}(\vec{q},\omega)\rangle \qquad (3)$$

A simplified dielectric response function of a metal may be written as

$$\varepsilon_m(q,\omega) = \left(1 - \frac{\omega_p^2}{\omega^2 - \omega_p^2 q^2/k^2}\right)\left(1 - \frac{\Omega_1^2(q)}{\omega^2}\right)\ldots\left(1 - \frac{\Omega_n^2(q)}{\omega^2}\right) \qquad (4)$$

where $\omega_p$ is the plasma frequency, $k$ is the inverse Thomas-Fermi radius, and $\Omega_n(q)$ are dispersion laws of various phonon modes [14]. Poles of the inverse dielectric response function $\varepsilon^{-1}{}_m(q,\omega)$ of metal (which correspond to various bosonic modes) maximize electron-electron pairing interaction given by Eq. (1).

Let us now consider the geometry of a fractal metal-dielectric metamaterial shown schematically in Fig. 1a. The effective dielectric response function of such a metamaterial may be obtained using the method of iterations. In the first iteration (the second panel in Fig. 1a) we will consider a metamaterial formed by mixing of a superconducting "matrix" with dielectric "inclusions" (described by the dielectric constants $\varepsilon_m$ and $\varepsilon_d$, respectively) where we will assume that the volume fractions of these components are $n$ and $(1 - n)$, respectively ($0 \leq n \leq 1$). In the particular case shown in Fig. 1a $n = 0.56$. According to the Maxwell-Garnett approximation [15], mixing of these components results in the effective medium with a dielectric constant $\varepsilon_{eff}$, which may be obtained as

$$\left(\frac{\varepsilon_{eff} - \varepsilon_m}{\varepsilon_{eff} + 2\varepsilon_m}\right) = (1-n)\left(\frac{\varepsilon_d - \varepsilon_m}{\varepsilon_d + 2\varepsilon_m}\right) \tag{5}$$

The explicit expression for $\varepsilon_{eff}$ may be written as

$$\varepsilon_{eff} = \frac{\varepsilon_m\left((3-2n)\varepsilon_d + 2n\varepsilon_m\right)}{\left(n\varepsilon_d + (3-n)\varepsilon_m\right)} \tag{6}$$

or

$$\varepsilon_{eff}^{-1} = \frac{n}{(3-2n)}\frac{1}{\varepsilon_m} + \frac{9(1-n)}{2n(3-2n)}\frac{1}{\left(\varepsilon_m + (3-2n)\varepsilon_d/2n\right)} \tag{7}$$

The superconducting critical temperature $T_c$ of such a "first fractal order" metamaterial has been calculated in [3] based on the consideration of the additional pole of the inverse dielectric response function, which is observed at

$$\varepsilon_m \approx -\frac{3-2n}{2n}\varepsilon_d \tag{8}$$

(see Eq. (7)). The additive contributions of this additional pole and the original metal pole ($\varepsilon_m \approx 0$) to the value of $\lambda_{\text{eff}}$ have been calculated in [3] as

$$\lambda_{\text{eff}} \approx \frac{9(1-n)}{2(3-2n)} \frac{\varepsilon_m''}{\varepsilon_{mm}''} \lambda_m = \frac{9(1-n)}{2(3-2n)} \alpha \lambda_m \qquad (9)$$

and

$$\lambda_{\text{eff}} \approx \frac{n^2}{(3-2n)} \lambda_m \qquad (10)$$

respectively, where $\varepsilon_m'' = \text{Im}\varepsilon_m$ at the original metal pole ($\varepsilon_m \approx 0$) value, and $\varepsilon_{mm}'' = \text{Im}\varepsilon_m$ at the additional pole described by Eq. (8). Since both poles are located not far from each other as a function of frequency $\omega$, the value of $\alpha = \varepsilon_m''/\varepsilon_{mm}''$ appears to be close to 1 [3]. Note that the additional "metamaterial" pole described by Eq. (8) disappears as $n \to 0$ at some critical value of the volume fraction $n = n_{\text{cr}}$ due to the finiteness of $\varepsilon_m$ (see Eq. (4)). Substitution of Eqs. (9, 10) into Eq. (2) produces the following final expression for the critical temperature of the "first fractal order" metamaterial [3]:

$$T_c = T_{Cbulk} \exp\left(\frac{1}{\lambda_m}\left(1 - \frac{1}{\left(\frac{n^2}{(3-2n)} + \frac{9(1-n)\alpha}{2(3-2n)}\right)}\right)\right) \quad \text{at } n > n_{\text{cr}}, \qquad (11)$$

Application of this model to the Al-Al$_2$O$_3$ core-shell metamaterial studied in [5] (assuming the known values $Tc_{\text{bulk}} = 1.2$ K, $\theta_D = 428$ K and $\lambda_m = 0.17$ of bulk aluminium [14]) resulted in excellent fit to the experimental data [3]. Such a good match unambiguously identifies the metamaterial enhancement as the physical mechanism of critical temperature tripling in the Al-Al$_2$O$_3$ core-shell metamaterial superconductors. Similar good agreement was also observed in the case of tin-BaTiO$_3$ ENZ metamaterial [3, 4].

The appearance of an additional "metamaterial" pole in the Maxwell-Garnett expression (Eq. (7)) for a metal-dielectric mixture does not rely on any particular spatial scale. The Maxwell-Garnett approximation for the dielectric constant of a mixture may be traced to the Clausius–Mossotti formula [15], which relates the dielectric constant of a mixture to polarizabilities and volume fractions of its constituents. Therefore, it should be applicable all the way down to the nanometer scale. Moreover, relatively large value of the superconducting coherence length $\xi = 1600$ nm of bulk aluminum [14] leaves a lot of room for engineering an aluminum-based fractal metamaterial superconductor according to the procedure shown schematically in Fig.1a.

As a next iteration, let us consider the central image (the 2nd fractal order) from Fig. 1a. Once again, we may use the Maxwell-Garnett approximation (Eq. (5)). However, this time we will assume that the "first order metamaterial" described by Eq. (6) plays the role of the "matrix". The volume fraction of the new matrix is $n$, and it is diluted with the same dielectric $\varepsilon_d$ with volume fraction $(1-n)$. Mixing of these components results in the effective medium with a dielectric constant $\varepsilon^{(2)}_{eff}$, which may be obtained as (compare to Eq. (6)):

$$\varepsilon^{(2)}_{eff} = \frac{\varepsilon^{(1)}_{eff}\left((3-2n)\varepsilon_d + 2n\varepsilon^{(1)}_{eff}\right)}{\left(n\varepsilon_d + (3-n)\varepsilon^{(1)}_{eff}\right)} \tag{12}$$

It is easy to verify that in addition to the pole at $\varepsilon_m \approx 0$ and the first order pole described by Eq. (8), the inverse dielectric response function of the "second order" fractal metamaterial has additional poles at

$$\varepsilon_m = -\varepsilon_d \frac{(3-2n)(3+n)}{8n^2}\left(1 \pm \sqrt{1 - \frac{16n^3}{(3-2n)(3+n)^2}}\right) \quad (13)$$

These additional poles may be expressed in the recurrent form as

$$\varepsilon_m^{(2)} = \varepsilon_m^{(1)} \frac{(3+n)}{4n}\left(1 \pm \sqrt{1 - \frac{16n^3}{(3-2n)(3+n)^2}}\right) \quad (14)$$

The latter equation, which expresses $\varepsilon_m^{(k+1)}$ as a function of $\varepsilon_m^{(k)}$ may be used to determine all the poles of the fractal metamaterial structure using an iterative procedure. The first few poles determined in such a way are plotted in Fig. 1b as a function of $n$ for the case of $\varepsilon_d = 3$. This plot makes it clear that most of the poles in each iteration (which correspond to the minus sign in Eq. (14)) stay near $\varepsilon_m \approx 0$, while one pole in each iteration (corresponding to the plus sign) may be written as

$$\varepsilon_m^{(k+1)} \approx \varepsilon_m^{(k)} \frac{(3+n)}{2n} \quad (15)$$

The vertical dashed line in Fig. 1b corresponds to the critical value of the metal volume fraction $n_{cr}$ observed in the case of the Al-Al$_2$O$_3$ core-shell superconductor [3]. As indicated in Fig. 1b, this value may be used as a guideline to evaluate how many additional poles may be observed in a real fractal metamaterial at a given volume fraction $n$. The additional poles of the inverse dielectric response function of the fractal metamaterial defined by Eq. (15) may be described as the plasmon-phonon modes of the metamaterial, which participate in the Cooper pairing of electrons in a fashion, which is quite similar to the conventional BCS mechanism.

Let us evaluate what kind of critical temperature increase may be expected in a fractal metamaterial superconductor because of the appearance of all these additional poles in $\varepsilon^{-1}_{\text{eff}}(q,\omega)$. We already have demonstrated that the additional poles of the inverse dielectric response function may be found using an iterative procedure (see Eqs. (14, 15)). A similar iterative expression may be derived for $\mathrm{Im}\varepsilon^{-1}_{\mathrm{eff}}(q,\omega)$ near these poles. Based on Eq. (7), we may write the following expression for the imaginary part of the inverse dielectric response function of the first order material near the metamaterial pole:

$$\mathrm{Im}\left(\left(\varepsilon_{\mathrm{eff}}^{(1)}\right)^{-1}\right) \approx \frac{9(1-n)}{2n(3-2n)} \alpha \, \mathrm{Im}\left(\left(\varepsilon_{\mathrm{eff}}^{(0)}\right)^{-1}\right), \quad (16)$$

where $\mathrm{Im}\left(\left(\varepsilon_{\mathrm{eff}}^{(0)}\right)^{-1}\right) = 1/\varepsilon_m''$, and $\alpha \sim 1$ near $n \sim 1$. Since the magnitude of $\varepsilon_m$ is finite (see Fig. 1b), in a real fractal metamaterial geometry a large number of poles may be expected only at values of $n$ close to 1. Therefore, from now on we will assume that $\alpha = 1$. An expression similar to Eq. (16) may be applied iteratively to the next fractal orders. As far as the density of states $\nu$ in Eq. (3) is concerned, as illustrated by Fig. 1a, with each iteration the density of states of the metamaterial is multiplied by a factor of $n$. Thus, we may now sum up contributions of all the fractal poles to $\lambda_{\mathrm{eff}}$. It may be obtained as a sum of a geometrical progression:

$$\lambda_{\mathrm{eff}} \approx \left( n^{K-1} \left( \frac{9(1-n)}{2(3-2n)} \right) + n^{K-2} \left( \frac{9(1-n)}{2(3-2n)} \right)^2 + \ldots + \left( \frac{9(1-n)}{2(3-2n)} \right)^K \right) \lambda_m =$$

$$= n^{K-1} \left( \frac{9(1-n)}{2(3-2n)} \right) \frac{\left(1 - \left(\frac{9(1-n)}{2n(3-2n)}\right)^K\right)}{\left(1 - \left(\frac{9(1-n)}{2n(3-2n)}\right)\right)} \lambda_m, \quad (17)$$

assuming that the metamaterial structure contains $K$ fractal orders. The resulting $\lambda_{\mathrm{eff}}$ is plotted in Fig. 2 for the first four fractal orders. The corresponding $n_{cr}^{(k)}$ for each fractal order (determined using Fig. 1b) are marked by the vertical dashed lines in Fig. 2. This figure clearly demonstrates that fractal structure promotes superconductivity by increasing $\lambda_{eff}$. For the case of Al-based metamaterial the enhancement of Tc is expected for at least the first three fractal orders. Compared to a simple metal-dielectric ENZ metamaterial (described as a "first order fractal" in Fig.2), the higher order fractal metamaterial structures exhibit much stronger enhancement of the coupling constant $\lambda_{eff}$, which should lead to much larger $T_c$ increase. Based on Fig. 2, the enhancement of $T_c$ ($\lambda_{eff}/\lambda_m > 1$) in a fractal metamaterial occurs starting at $n \sim 0.8$, which is close enough to $n \sim 1$ to validate our assumptions. The corresponding values of $T_c$ calculated as

$$T_c = T_{Cbulk} \exp\left(\frac{1}{\lambda_m} - \frac{1}{\lambda_{eff}}\right) \quad (18)$$

based on the superconducting parameters of bulk aluminium for the first two fractal orders are shown in Fig. 3. The calculated curves are terminated at the experimentally defined values of $n_{cr}^{(k)}$ for the respective fractal order, which are taken from Fig. 1b. Eight-fold enhancement of $T_c$ compared to the bulk aluminium is projected for the second order fractal structure. The fractal enhancement of Tc may become even more pronounced for materials having larger values of $-\varepsilon_m$ compared to bulk aluminum.

Our results agree with the recent observations [10, 11] that fractal defect structure promotes superconductivity. In particular, it was observed that the microstructures of the transition-metal oxides, including high-$T_c$ copper oxide superconductors, are complex. For example, the oxygen interstitials or vacancies (which strongly influence the dielectric properties of the bulk high-$T_c$ superconductors [16]) exhibit fractal order [10]. These oxygen interstitials are located in the spacer layers separating the superconducting $CuO_2$ planes. They undergo fractal ordering phenomena that induce enhancements in the transition temperatures with no changes in the overall hole concentrations. Such ordering of oxygen interstitials in the $La_2O_{2+y}$ spacer layers of $La_2CuO_{4+y}$ high-$T_c$ superconductors is characterized by a fractal distribution up to a maximum limiting size of 400 mm. It was observed [10], quite intriguingly, that these fractal distributions of dopants seem to enhance superconductivity at high temperature, which is difficult to explain since the superconducting coherence length in these compounds is very small $\xi \sim 1 - 2$ nm [10, 11]. Appearance of the plasmon-phonon modes of the fractal structure described above may resolve this issue, since such fractal phonon modes reside on a scale, which is much larger than several nanometers. We should also mention that such a mechanism may also explain observations of increased critical temperatures in fractal Pb thin films [17].

In conclusion, we have proposed a fractal metamaterial superconductor geometry, which has been analyzed based on the recently developed theoretical description of critical temperature increase in epsilon near zero (ENZ) metamaterial superconductors. Considerable enhancement of critical temperature has been predicted in such materials due to appearance of large number of additional poles in the inverse dielectric response function of the fractal, which correspond to plasmon-phonon modes of the metamaterial. Our results agree with the recent observations [10, 11,17] that fractal defect structure promotes superconductivity.

This work was supported in part by NSF grant DMR-1104676.

**Additional Information**

Competing financial interests.

The authors declare no competing financial interests.

**Figure Captions**

**Figure 1**. (a) Schematic geometry of a metal-dielectric fractal metamaterial superconductor. For the first fractal order (the secondpanel) the volume fraction of the superconductor is $n = 0.56$. (b) The first few poles of the inverse dielectric response function of a fractal metal-dielectric metamaterial plotted as a function of $n$ for the case of $\varepsilon_d = 3$. The fractal order is marked near the curve for each pole. The vertical dashed line corresponds to the critical value of the metal volume fraction $n_{cr}$ observed in the case of the Al-Al$_2$O$_3$ core-shell superconductor. It is used to determine $n_{cr}^{(k)}$ for the next fractal orders, as indicated by the horizontal dashed line.

**Figure 2**. Plot of the magnitude of $\lambda_{\text{eff}}$ given by Eq. (17) as a function of $n$ for the first four orders of a fractal metamaterial superconductor. The corresponding $n_{cr}^{(k)}$ are marked by the vertical dashed lines.

**Figure 3**. Enhancement of $T_c$ of the fractal metamaterial superconductor calculated as a function of $n$ based on the superconducting parameters of bulk aluminium. The second order fractal structure demonstrates considerably higher $T_c$ compared to the first order structure.

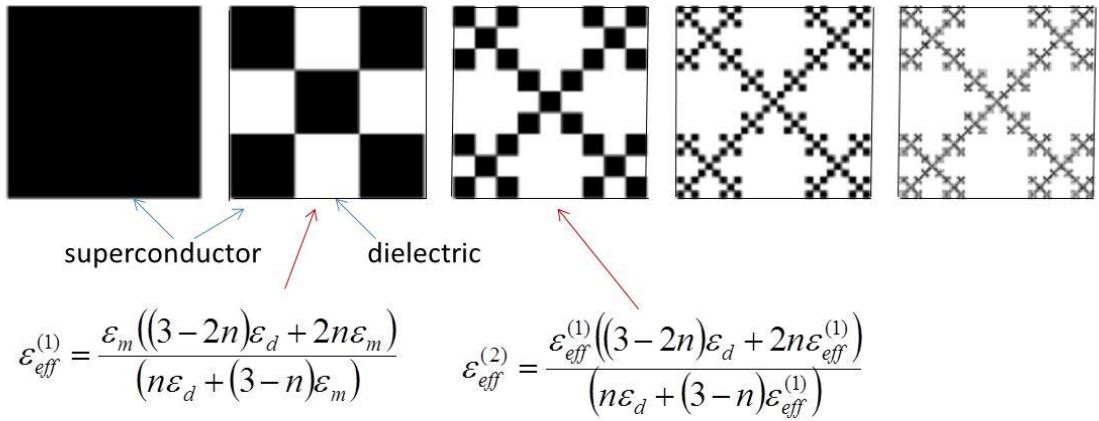

(a)

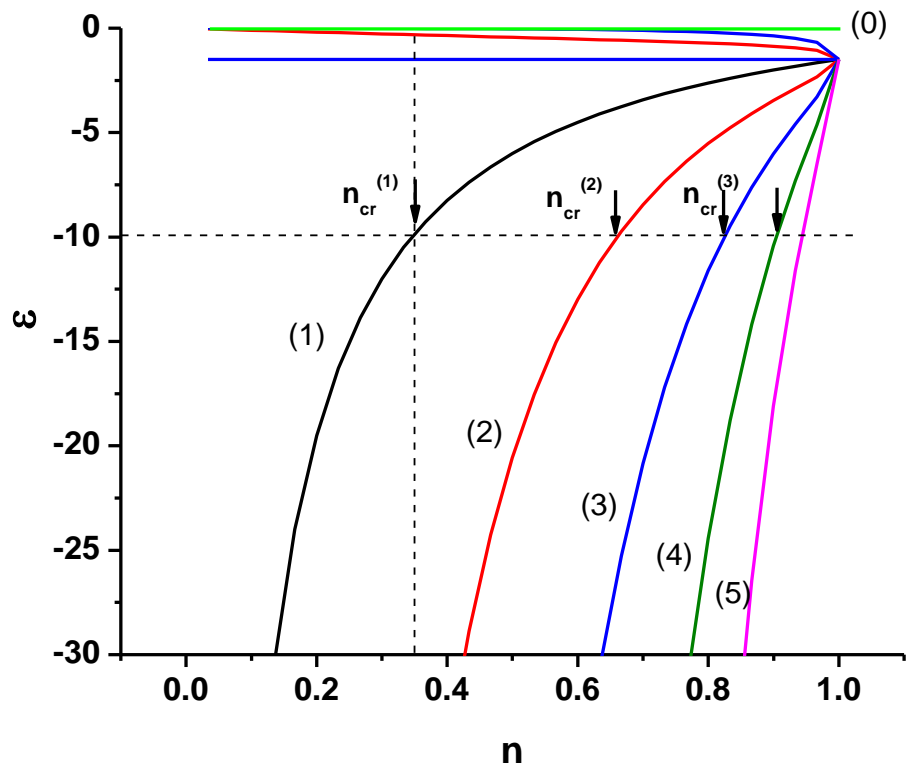

(b)

Fig. 1

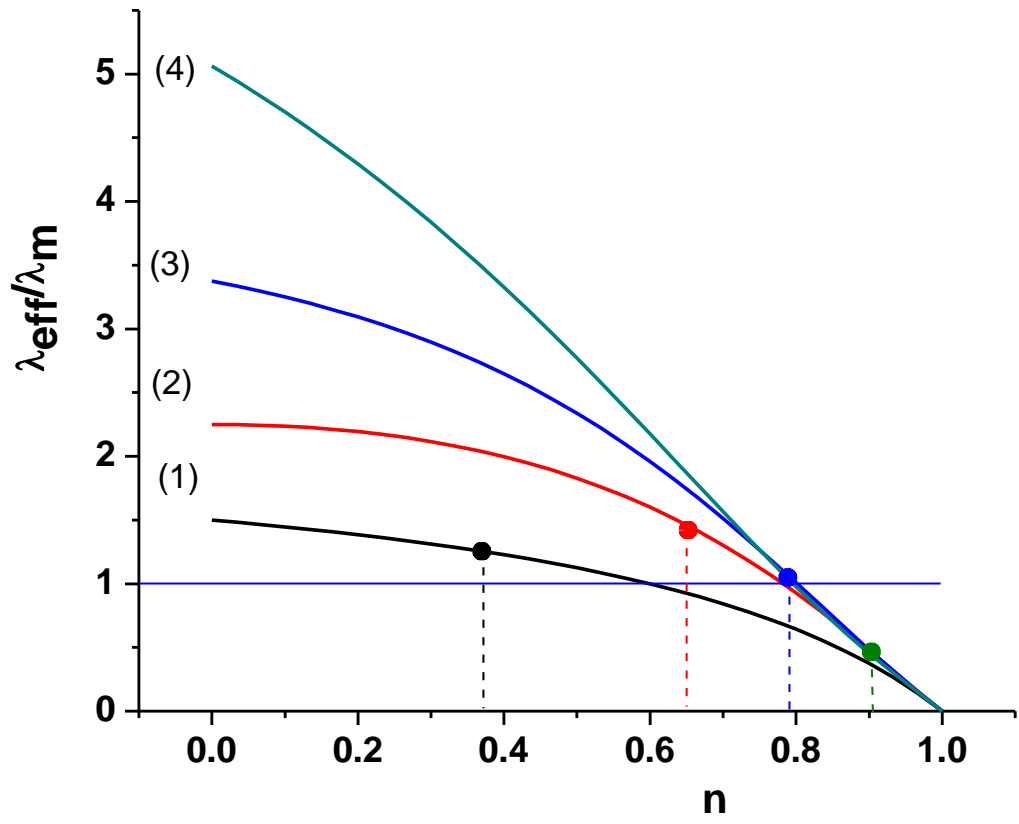

Fig. 2

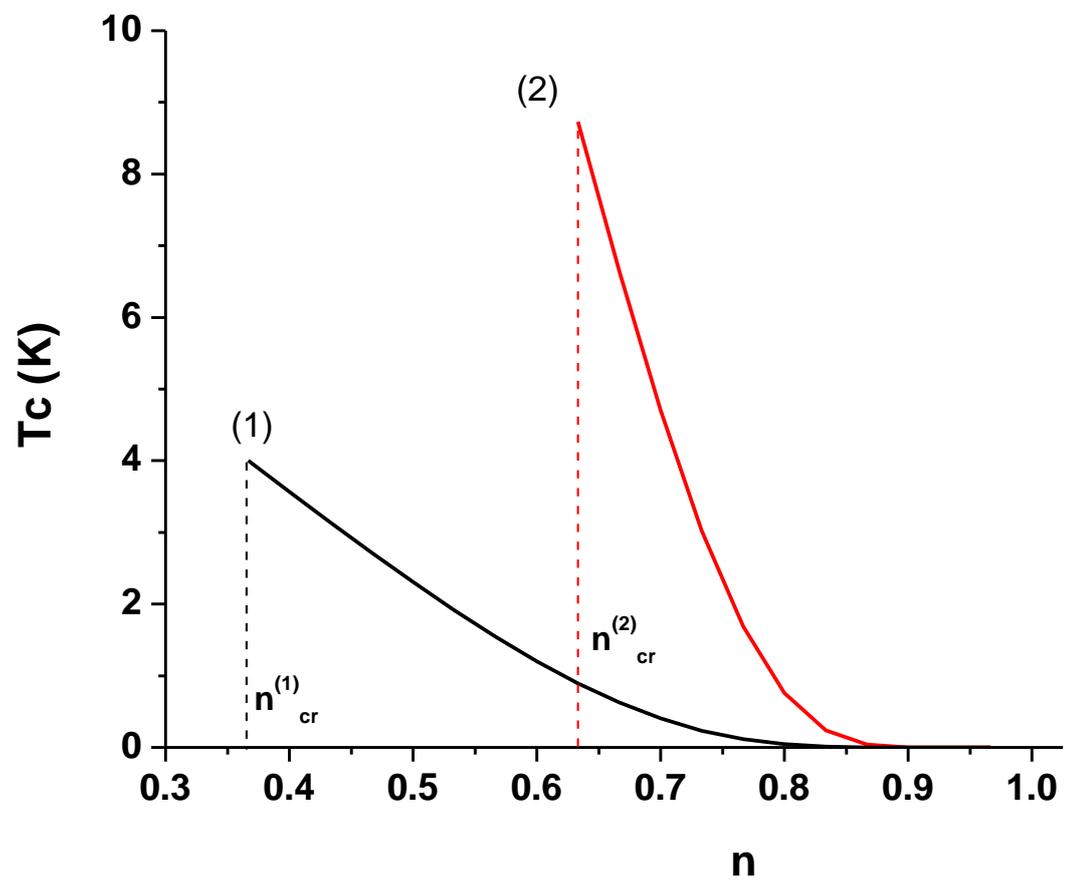

Fig. 3